\documentclass[aps,prl,preprint,tightenlines,superscriptaddress,showpacs,byrevtex]{revtex4}
\usepackage{graphicx}
\usepackage{epsfig}
\usepackage{dcolumn}

\newcommand{\MMS}{M_{\rm rec}^2}
\newcommand{\y}{Y(2175)}
\newcommand{\lum}{{\cal L}}
\newcommand{\eff}{\varepsilon}
\newcommand{\BR}{{\cal B}}
\newcommand{\jpc}{J^{PC}}
\newcommand{\pip}{\pi^+}
\newcommand{\pim}{\pi^-}
\newcommand{\piz}{\pi^0}
\newcommand{\kap}{K^+}
\newcommand{\kam}{K^-}

\newcommand{\jpsi}{J/\psi}
\newcommand{\EE}{e^+e^-}

\newcommand{\pp}{\pi^+\pi^-}
\newcommand{\fzero}{f_0(980)}
\newcommand{\kk}{K^+K^-}

\newcommand{\beq}{\begin{equation}}
\newcommand{\eeq}{\end{equation}}
\newcommand{\beqy}{\begin{eqnarray}}
\newcommand{\eeqy}{\end{eqnarray}}
\newcommand{\bitm}{\begin{itemize}}
\newcommand{\eitm}{\end{itemize}}

\begin{document}

\preprint{\vbox{ \hbox{   }
                 \hbox {  }
                        \hbox{Belle Preprint 2009-12}
                        \hbox{KEK   Preprint 2009-8}
                        \hbox{BIHEP-EP-2009-1}
                         }}
\title{\quad\\[1.5cm]
Observation of the $\phi(1680)$ and the $\y$ in $\EE \to \phi\pp$}

\affiliation{Budker Institute of Nuclear Physics, Novosibirsk}
\affiliation{Chiba University, Chiba} \affiliation{University of
Cincinnati, Cincinnati, Ohio 45221} \affiliation{T. Ko\'{s}ciuszko
Cracow University of Technology, Krakow}
\affiliation{Justus-Liebig-Universit\"at Gie\ss{}en, Gie\ss{}en}
\affiliation{The Graduate University for Advanced Studies, Hayama}
\affiliation{Gyeongsang National University, Chinju}
\affiliation{Hanyang University, Seoul} \affiliation{University of
Hawaii, Honolulu, Hawaii 96822} \affiliation{High Energy Accelerator
Research Organization (KEK), Tsukuba} \affiliation{Hiroshima
Institute of Technology, Hiroshima} \affiliation{University of
Illinois at Urbana-Champaign, Urbana, Illinois 61801}
\affiliation{Institute of High Energy Physics, Chinese Academy of
Sciences, Beijing} \affiliation{Institute of High Energy Physics,
Vienna} \affiliation{Institute of High Energy Physics, Protvino}
\affiliation{Institute for Theoretical and Experimental Physics,
Moscow} \affiliation{J. Stefan Institute, Ljubljana}
\affiliation{Kanagawa University, Yokohama} \affiliation{Institut
f\"ur Experimentelle Kernphysik, Universit\"at Karlsruhe, Karlsruhe}
\affiliation{Korea University, Seoul}
\affiliation{Kyungpook National University, Taegu}
\affiliation{\'Ecole Polytechnique F\'ed\'erale de Lausanne (EPFL),
Lausanne} \affiliation{Faculty of Mathematics and Physics,
University of Ljubljana, Ljubljana} \affiliation{University of
Maribor, Maribor} \affiliation{University of Melbourne, School of
Physics, Victoria 3010} \affiliation{Nagoya University, Nagoya}
\affiliation{Nara Women's University, Nara} \affiliation{National
Central University, Chung-li} \affiliation{National United
University, Miao Li} \affiliation{Department of Physics, National
Taiwan University, Taipei} \affiliation{H. Niewodniczanski Institute
of Nuclear Physics, Krakow} \affiliation{Nippon Dental University,
Niigata} \affiliation{Niigata University, Niigata}
\affiliation{University of Nova Gorica, Nova Gorica}
\affiliation{Novosibirsk State University, Novosibirsk}
\affiliation{Osaka City University, Osaka}
\affiliation{Panjab University, Chandigarh}
\affiliation{Saga University, Saga} \affiliation{University of
Science and Technology of China, Hefei} \affiliation{Seoul National
University, Seoul}
\affiliation{Sungkyunkwan University, Suwon} \affiliation{University
of Sydney, Sydney, New South Wales}
\affiliation{Toho University, Funabashi} \affiliation{Tohoku Gakuin
University, Tagajo} \affiliation{Tohoku University, Sendai}
\affiliation{Department of Physics, University of Tokyo, Tokyo}
\affiliation{Tokyo Metropolitan University, Tokyo}
\affiliation{Tokyo University of Agriculture and Technology, Tokyo}
\affiliation{IPNAS, Virginia Polytechnic Institute and State
University, Blacksburg, Virginia 24061} \affiliation{Yonsei
University, Seoul}
  \author{C.~P.~Shen}\affiliation{Institute of High Energy Physics, Chinese Academy of Sciences, Beijing}\affiliation{University of Hawaii, Honolulu, Hawaii 96822} 
  \author{C.~Z.~Yuan}\affiliation{Institute of High Energy Physics, Chinese Academy of Sciences, Beijing} 
  \author{P.~Wang}\affiliation{Institute of High Energy Physics, Chinese Academy of Sciences, Beijing} 
  \author{X.~L.~Wang}\affiliation{Institute of High Energy Physics, Chinese Academy of Sciences, Beijing} 
  \author{I.~Adachi}\affiliation{High Energy Accelerator Research Organization (KEK), Tsukuba} 
  \author{H.~Aihara}\affiliation{Department of Physics, University of Tokyo, Tokyo} 
  \author{K.~Arinstein}\affiliation{Budker Institute of Nuclear Physics, Novosibirsk}\affiliation{Novosibirsk State University, Novosibirsk} 
  \author{V.~Aulchenko}\affiliation{Budker Institute of Nuclear Physics, Novosibirsk}\affiliation{Novosibirsk State University, Novosibirsk} 
  \author{A.~M.~Bakich}\affiliation{University of Sydney, Sydney, New South Wales} 
  \author{E.~Barberio}\affiliation{University of Melbourne, School of Physics, Victoria 3010} 
% ...
\collaboration{The Belle Collaboration}

\noaffiliation
\date{\today}
\begin{abstract}

The cross sections for $e^+ e^- \to \phi\pi^+\pi^-$ and $e^+ e^- \to
\phi \fzero$ are measured from threshold to
$\sqrt{s}=3.0$~$\hbox{GeV}$ using initial state radiation. The
analysis is based on a data sample of 673 fb$^{-1}$ collected on and
below the $\Upsilon(4S)$ resonance with the Belle detector at the
KEKB asymmetric-energy $e^+e^-$ collider. First measurements are
reported for the resonance parameters of the $\phi(1680)$ in the
$\phi\pi^+\pi^-$ mode: $m=(1689\pm 7\pm 10)$~MeV/$c^2$ and
$\Gamma=(211\pm 14\pm 19)$~MeV/$c^2$. A structure at
$\sqrt{s}=2.1~\hbox{GeV}/c^2$, corresponding to the so called
$Y(2175)$, is observed; its mass and width are determined to be
$2079\pm13^{+79}_{-28}$~MeV/$c^2$ and
$192\pm23^{+25}_{-61}~\hbox{MeV}/c^2$, respectively.

\end{abstract}

\pacs{13.66.Bc, 13.25.-k, 14.40.Cs}

\maketitle

Although vector mesons are produced copiously in $\EE$ annihilation,
the resonance parameters of excited vector states are not well
measured~\cite{PDG}. In the $s\bar{s}$ sector, the $\phi(1680)$ was
first observed by the DM1 Collaboration in the reaction $\EE \to K
\bar{K} \pi$~\cite{DM1} and was recently studied in the
initial-state radiation (ISR) events of the $\EE \to \phi \eta$ and
$K^{\ast}(892) \bar{K}+c.c.$ modes by the BaBar Collaboration, which
measured its mass and width to be $1723\pm 20$~$\hbox{MeV}/c^2$ and
$371\pm 75$~$\hbox{MeV}/c^2$, respectively~\cite{babar1680}. In a
study of ISR events of the type, $\EE \to \gamma_{\rm ISR} \phi
\pp$, the BaBar Collaboration observed two clear structures near
$\sqrt{s}=1.68$~GeV/$c^2$ and 2.175~GeV/$c^2$, the latter was
produced dominantly via a $\phi \fzero$ intermediate state, and was
dubbed the $\y$~\cite{babar2175}. The BES Collaboration confirmed
the $\y$ in the $\phi\fzero$ invariant mass spectrum of $\jpsi \to
\eta \phi \fzero$ ($\eta \to \gamma \gamma$, $\phi \to \kap \kam$
and $\fzero \to \pip \pim$) decays with a statistical significance
of about 5$\sigma$~\cite{bes2175}. The Particle Data Group (PDG)
assigns all these observations to a new state referred to as the
$\phi(2170)$~\cite{PDG}.

Since the $\y$ resonance is produced via ISR in $\EE$ collisions,
its $\jpc=1^{--}$. This observation stimulated the theoretical
speculation that $\y$ may be an $s$-quark counterpart  of the
$Y(4260)$~\cite{babay4260,bn978} since both are produced in $\EE$
annihilation and exhibit similar decay patterns. On the other hand,
a number of different interpretations have been proposed for the
$\y$ with predicted masses that are consistent, within errors, with
the experimental measurements. These include: an $s\bar{s}g$
hybrid~\cite{ssg}; a $2^3D_1$ $s\bar{s}$ state~\cite{2D} with a
width predicted to be in the range 120-210~$\hbox{MeV}/c^2$; a
tetraquark state~\cite{4s,chenhx}; a $\Lambda \bar{\Lambda}$ bound
state~\cite{eberhard}; or an ordinary $\phi\fzero$ resonance
produced by interactions between the final state
particles~\cite{27mev}. The possibility that the $\y$ is a $3^3S_1$
$s\bar{s}$ state is disfavored by the rather large predicted width
($\Gamma \sim 380~\hbox{MeV}/c^2$)~\cite{3s}, which disagrees with
the experimental upper limit of 100~MeV/$c^2$~\cite{babar2175}. A
recent review~\cite{zhusl} discusses the basic problem of the large
expected decay widths into two mesons, which contrasts with
experimental observations.

In the analysis reported here, we use a data sample with an
integrated luminosity of 673~fb$^{-1}$ collected with the Belle
detector~\cite{Belle} operating at the KEKB asymmetric-energy
$e^+e^-$ (3.5 on 8~GeV) collider~\cite{KEKB} to investigate the
$\phi \pp$ final state via ISR. About 90\% of data were collected at
the $\Upsilon(4S)$ resonance ($\sqrt{s}=10.58~ \hbox{GeV}$), and the
rest were taken at a center-of-mass (C.M.) energy that is 60~MeV
below the $\Upsilon(4S)$ peak.

For Monte Carlo (MC) simulation of the ISR process, we generate
signal events with the PHOKHARA program~\cite{phokhara}. In this
program, after one or two photons are emitted, the lower energy
$\EE$ pair forms a resonance $X$ that subsequently decays to $\phi
\pp$ or $\phi \fzero$ with the $\phi$ decaying into $\kk$ and the
$\fzero$ decaying into $\pp$. In the $X\to \phi \pp$ generation, we
assume that the $\pi\pi$ system is pure $S$-wave and that it and the
$\phi$ are also in a relative $S$-wave. The $\pp$ invariant mass
distributions are generated according to phase space.

To select the $\phi \pp$ final states, we use the following event
selection criteria. We require four good charged tracks with net
charge equal to zero. Good tracks should have impact parameters
perpendicular to and along the beam direction with respect to the
interaction point of less than $0.3$ and $2$~cm, respectively, and
transverse momentum is restricted to be higher than 0.1~GeV/$c$. For
each charged track, the particle identification information from
different detector subsystems is combined to form a likelihood for
each particle species ($i$), $L_i$~\cite{pid}. Tracks with
$\mathcal{R}_K=\frac{L_K}{L_K+L_\pi}<0.9$ are identified as pions,
tracks with $\mathcal{R}_K>0.6$ are identified as kaons. We require
that two tracks be identified as pions, and assign kaon masses to
the other two tracks, of which only one is required to be identified
as a kaon. The sum of the charged track energies and that of
reconstructed photons (neutral clusters in the electromagnetic
calorimeter with energies greater than 100 MeV and not associated
with charged tracks in the central drift chamber) must be greater
than 9~GeV; this ensures that an ISR photon is detected. The
efficiency of this requirement is nearly 100\% for $\phi \pp$
invariant masses below 3.0~GeV/$c^2$. We require
$-1.0~(\hbox{GeV}/c^2)^2 <\MMS<2.0~(\hbox{GeV}/c^2)^2$ to remove
multi-photon backgrounds, where $\MMS$ is the square of the mass
that is recoiling against the four charged tracks. The MC simulation
indicates that the application of these requirements reduces the
contamination from the background ISR process $\EE \to \phi \pp
\pi^0 (\gamma)$~\cite{phieta} to the 1\% level; the effect of the
residual background is included as a systematic error.

\begin{figure}[htbp]
\centerline{\psfig{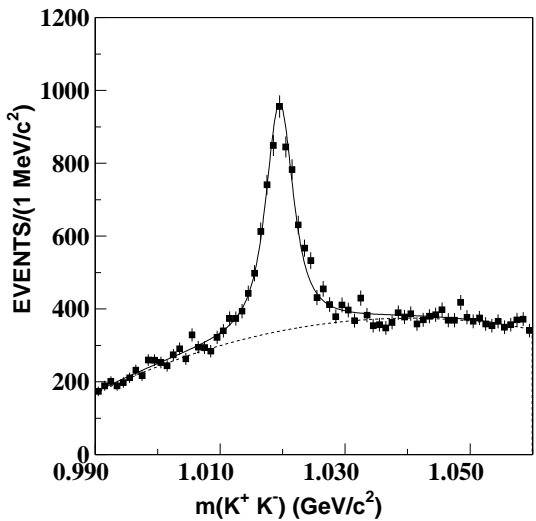}} \caption{ The $\kk$
invariant mass distribution for selected $\EE \to \kk \pp$
candidates with $\kk\pp$ invariant mass less than 3.0~GeV/$c^2$. The
curves show the fit projections for the full fit and for the
background component.} \label{mkk}
\end{figure}

In events with $\kk\pp$ mass below 3.0~$\hbox{GeV}/c^2$, a $\phi$
meson signal is evident in the $\kk$ invariant mass distribution, as
shown in Fig.~\ref{mkk}. This distribution is fitted with a
Breit-Wigner (BW) (the $\phi$ meson mass and width fixed to PDG
values) convolved with a Gaussian resolution function as the $\phi$
signal and a second-order polynomial as the background shape; here
the width of the Gaussian resolution function is fixed at
$\sigma=0.95$~MeV/$c^2$, its MC-determined value. The fit yields
$4832\pm 132$ $\phi\pp$ events.

We obtain the number of $\phi \pp$ events in each $\kk \pp$
invariant mass bin by fitting the $\kk$ invariant mass distribution
for each interval, using the same fit model as in the fit for the
overall sample.  In order to reliably control the background shape,
we restrict the coefficients of the background polynomials in nearby
bins to vary smoothly along parabolas. The parameters of these
parabolas are determined from fits to the coefficients obtained from
fits to the $\kk$ invariant mass distribution in each $\kk\pp$
invariant mass bin. The resulting $\phi \pp$ invariant mass
distribution is shown in Fig.~\ref{m2k2pi-fit}(a). Combinatorial
background has a smooth distribution in the $\kk$ invariant mass
distribution and does not affect the results since we fit the $\kk$
distribution to obtain the number of $\phi \pp$ events in each $\kk
\pp$ mass bin. In Fig.~\ref{m2k2pi-fit}(a) there are two distinct
peaks: one near 1.7~GeV/$c^2$ and another near 2.1~GeV/$c^2$,
corresponding to the $\phi(1680)$ and $\y$ states. In addition,
there is a cluster of events near 2.4~GeV/$c^2$.

\begin{figure}[htbp]
\psfig{file=fig2a.epsi, height=5 cm}\hspace{1cm}
\psfig{file=fig2b.epsi, height=5 cm} \caption{The $\phi \pp$
invariant mass distribution obtained from fitting to $m(\kk)$ in
each $m(\kk\pp)$ bin (a) as described in the text, and the measured
$\EE\to \phi \pp$ cross section (b). The errors are statistical
only.} \label{m2k2pi-fit}
\end{figure}

The $\EE\to \phi\pp$ cross section for each $\phi\pp$ mass bin is
computed using
 \[
 \sigma_i = \frac{n^{\rm fit}_i}
                 {\eff_i \lum_i \BR(\phi\to \kk)},
 \]
where $n^{\rm fit}_i$, $\eff_i$, and $\lum_i$ are the number of
$\phi\pp$ events fitted in data, the efficiency, and the effective
luminosity~\cite{kuraev} in the $i$-th $\phi \pp$ mass bin,
respectively; $\BR(\phi \to \kk)=49.2\%$~\cite{PDG}. According to
the MC simulation, the efficiency increases smoothly from 1.72\% to
2.67\% for $\phi \pp$ masses ranging from threshold to
3.0~GeV/$c^2$. The resulting cross sections are shown in
Fig.~\ref{m2k2pi-fit}(b)~\cite{database}.

\begin{figure}[htbp]
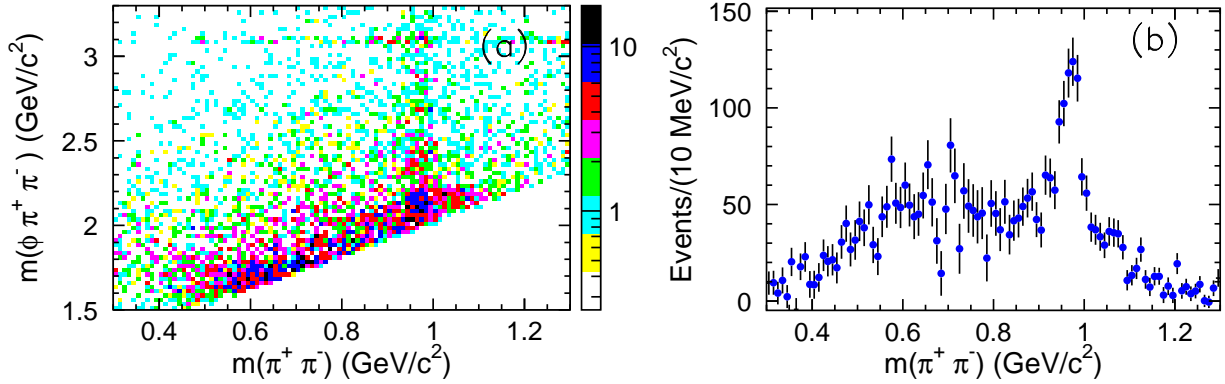

\psfig{file=fig3a.epsi,height=5cm} \hspace{0.1cm}
\psfig{file=fig3b.epsi,height=5cm} \caption{Scatter plot of
$m(\phi\pp)$ versus $m(\pp)$, where non-$\phi$ background is
subtracted according to the normalized $\phi$ mass sidebands (a),
and the projection on $\pp$ mass, a clear $\fzero$ signal is visible
(b).} \label{mpipi}
\end{figure}

We now consider the quasi-two-body intermediate state $\phi \fzero$.
Figure~\ref{mpipi}(a) shows a scatter plot of $m(\phi\pp)$ versus
$m(\pp)$ for events in the $\phi$ signal region ($1.013~{\rm
GeV}/c^2 < m_{\kk} < 1.025~{\rm GeV}/c^2$). Non-$\phi$ background is
subtracted according to the normalized $\phi$ mass sidebands
($m_{\kk}\in [0.989, 1.013]$~GeV/$c^2$ or $m_{\kk}\in [1.025,
1.049]$~GeV/$c^2$). It can be seen that within the $\fzero$ mass
band there are clusters of events that correspond to $\y\to
\phi\fzero$ and $\jpsi\to \phi\fzero$. Figure~\ref{mpipi}(b) shows
the projected $m(\pp)$ distribution, where a clear $\fzero$ signal
is evident. For each $25~\hbox{MeV}/c^2$ bin of $m(\kk \pp)$, we
select events with $m(\pp)\in
[0.85,1.10]~$\hbox{GeV}/$c^2$~\cite{footnote}, and fit their
$m(\kk)$ distribution to extract the number of $\phi \fzero$ events
in a way similar to that used to extract the number of $\phi\pp$
events described above. The resulting $\phi \fzero$ invariant mass
distribution is shown in Fig.~\ref{mphif0-fit}(a). A very clear $\y$
signal is observed with an accumulation of events around
2.4~GeV/$c^2$. The $\EE\to \phi\fzero$ cross section for each
$\phi\fzero$ mass bin is computed in the same way as that for the
$\phi\pp$ mode using $\BR(\fzero \to \pp)=\frac{2}{3}$. The
efficiency is nearly independent of $\phi \fzero$ mass, 2.3\% from
threshold to 3.0~GeV/$c^2$. The resulting cross sections are shown
in Fig.~\ref{mphif0-fit}(b).

\begin{figure}[htbp]
\psfig{file=fig4a.epsi,height=5cm} \hspace{1cm}
\psfig{file=fig4b.epsi,height=5cm}
 \caption{The $\phi \fzero$
invariant mass distribution (a) obtained from fitting to $m(\kk)$
in each $m(\kk\fzero)$ bin as described in the text, and (b) the
measured $\EE\to \phi \fzero$ cross sections. The errors are
statistical only.} \label{mphif0-fit}
\end{figure}

The sources of the systematic errors for the cross section
measurements are summarized in Table~\ref{err_full}. The uncertainty
is 1.0\% for pion identification, and is negligible for kaons since
the identification of only one of the kaons is required; the
uncertainty in the tracking efficiency is 1\% per track, and is
additive; the efficiency uncertainty associated with the $\MMS$
requirement is determined to be 1.0\%~\cite{bn978}; the
uncertainties in the yields of $\phi \pp$ and $\phi \fzero$ events
for each mass bin due to the $\phi$ signal fit are estimated to be
$5.0\%$ and $4.0\%$ as determined by: changing the $\phi$ mass
resolution, the orders of the background polynomial, and the fitting
range. The error due to the uncertainty of the background parameters
changes with $m_{\phi \pp}$~(GeV/$c^2$) because of signal-to-noise
ratio variations between threshold and 3~GeV/$c^2$. The error is
approximated as $(5.92-5.44 m_{\phi \pp}+1.28 m_{\phi
\pp}^2)/(0.02+(m_{\phi \pp}-1.9)^2)\%$, $m_{\phi\pp}$ in~GeV/$c^2$,
which is about 4\% in the $\phi(1680)$ and the $\y$ mass regions,
and increases continuously to about 10\% in between and decreases to
about 1\% above 2.3~GeV/$c^2$. This is common to the $\phi\pp$ and
$\phi\fzero$ modes. The $\phi\pp\piz(\gamma)$ background further
contributes a 1\% uncertainty to the $\phi\pp$, and a negligible
fraction to the $\phi\fzero$ mode. Belle measures luminosity with a
precision of 1.4\% using wide angle Bhabha events. The uncertainty
of the ISR photon radiator is 0.1\%~\cite{kuraev}. The main
uncertainty in the PHOKHARA~\cite{phokhara} generator is due to the
modelling of the $\pp$ mass spectrum. This is tested by generating
events with different $\pp$ mass distributions. We take 5\% and 3\%
as conservative uncertainties related to the MC generators for the
$\phi \pp$ and $\phi \fzero$ modes, respectively. According to the
MC simulation, the trigger efficiency for these events is around
$(97\pm1)\%$ with little dependence on the $\phi \pp$ invariant
mass. The uncertainty of $\BR(\phi\to \kk)$ is 1.2\%~\cite{PDG}.
Finally the error due to MC statistics is 0.8\%. We assume all the
sources are independent and add them (excluding that from background
parameters) in quadrature, resulting in the total systematic errors
on the cross sections of 8.6\% and 6.9\% for $\phi\pp$ and
$\phi\fzero$ modes, respectively.

\begin{table}[htbp]
\caption{Systematic errors (\%) in the $\phi \pp$ and $\phi \fzero$
cross sections measurements. The systematic error from the
uncertainty of background parameters [$(5.92-5.44 m_{\phi \pp}+1.28
m_{\phi \pp}^2)/(0.02+(m_{\phi \pp}-1.9)^2)\%$, $m_{\phi \pp}$ in
~GeV/$c^2$] is not included. } \label{err_full}
\begin{center}
\begin{tabular}{c  c  c }
\hline
  Source &  ~~$\phi\pp$~~&  ~~$\phi \fzero$~~ \\\hline
 Particle ID&  1.0 & 1.0 \\
 Tracking & 4.0 & 4.0 \\
 $\MMS$ selection & 1.0 & 1.0  \\
 $\phi$ signal fit & 5.0 & 4.0 \\
 $\phi\pp\piz(\gamma)$ background & 1.0 & 0.0\\
 Integrated luminosity & 1.4 & 1.4  \\
 Generator & 5.0 & 3.0 \\
 Trigger efficiency & 1.0 & 1.0\\
 Branching fractions & 1.2 &1.2 \\
 MC statistics & 0.8 &0.8 \\
 \hline
 Sum in quadrature & 8.6 & 6.9\\
 \hline
\end{tabular}
\end{center}
\end{table}

\begin{figure}[htbp]
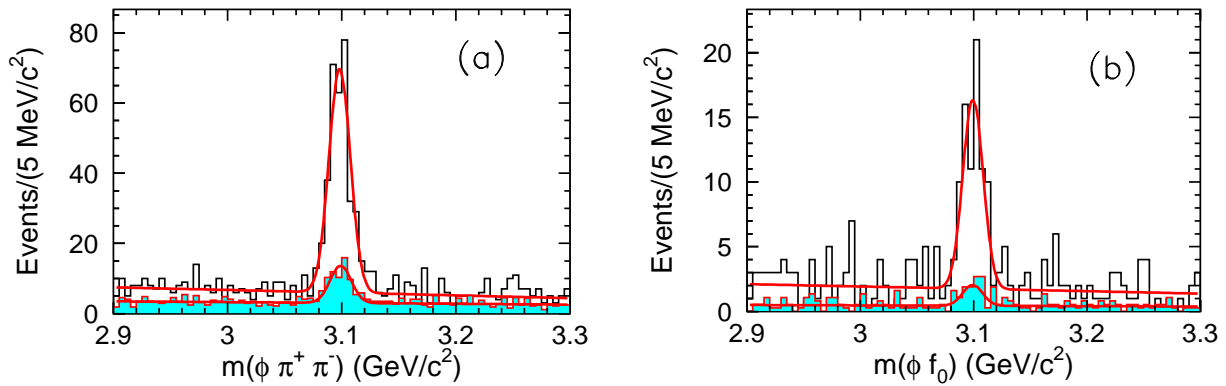

\psfig{file=fig5a.epsi,height=5cm} \hspace{0.5cm}
\psfig{file=fig5b.epsi,height=5cm} \caption{The $\jpsi\to \phi\pp$
(a) and $\jpsi\to \phi \fzero$ (b) signals in the $\phi$ signals
regions (open histograms) and in the $\phi$ sidebands (shaded
histograms, normalized). The curves are the best fits to the signal
and sidebands distributions, respectively.} \label{jpsi}
\end{figure}

There are strong $\jpsi$ signals in both $\phi \pp$ and $\phi
\fzero$ samples (see Fig.~\ref{jpsi}). $254\pm 23$ $\jpsi \to \phi
\pp$ and $60\pm 11$ $\jpsi \to \phi \fzero$ events are obtained from
fitting the $\jpsi$ signals and subtracting the peaking backgrounds
by fitting the normalized $\phi$ sidebands events. We determine
$\BR(\jpsi \to \phi \pp)\times \Gamma_{\EE}=(4.50\pm 0.41\pm
0.26)$~eV/$c^2$ and $\BR(\jpsi \to \phi \fzero)\times \Gamma_{\EE}=
(2.22\pm 0.41\pm 0.13)$~eV/$c^2$, where the first errors are
statistical, and the second systematic. The sources of the
systematic errors are almost the same as those listed in
Table~\ref{err_full}, except that the uncertainty of the $\phi$
signal fit is replaced by that of the $\jpsi$ signal fit. Our
results are consistent with the PDG values~\cite{PDG} and BaBar's
measurements~\cite{babar2175}, but are more precise.

\begin{figure}[htbp]
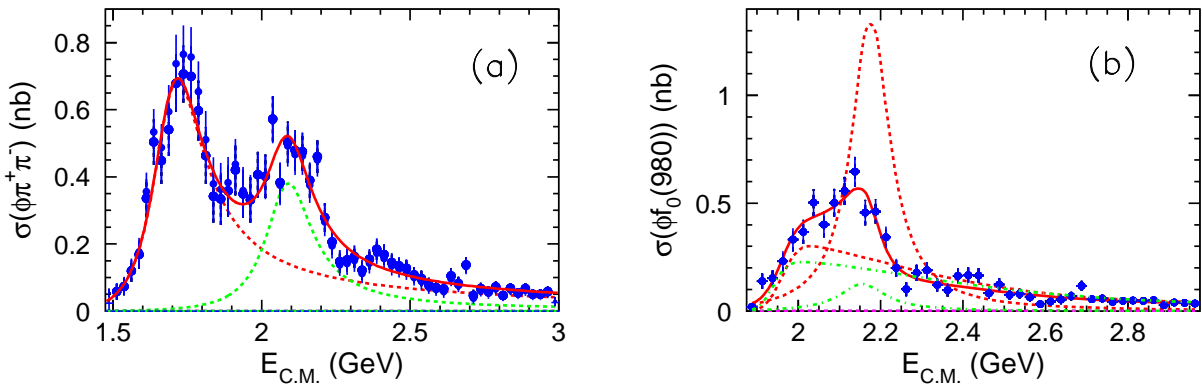

\psfig{file=fig6a.epsi,height=5cm}\hspace{1cm}
\psfig{file=fig6b.epsi,height=5cm} \caption{Fit to (a) $\EE\to \phi
\pp$ cross section with two incoherent BW functions, one for the
$\phi(1680)$ and the other for the $\y$ and (b) $\EE\to \phi \fzero$
cross section with a single BW function that interferes with a
non-resonant component. The curves show the projections from the
best fit and the contribution from each component. In (b), the
dashed curves are for the destructive interference solution and the
dot-dashed curves for the constructive interference solution}.
\label{belle-oneres}
\end{figure}

In order to obtain the resonance parameters of the $\phi(1680)$ and
$\y$, a least squares fit is applied to the $\phi \pp$ cross section
distribution. Since the $\phi(1680)$ decays into $\phi \pp$ while
the $\y$ decays dominantly into $\phi \fzero$, we use two incoherent
BW functions in the fit, one for the $\phi(1680)$ and the other for
the $\y$. The fit result is shown in Fig.~\ref{belle-oneres}(a),
with a goodness-of-the-fit of $\chi^2/ndf=68/55$, corresponding to a
C.L. of 12\%. The statistical significance of each resonance is
greater than 10$\sigma$. From the fit we obtain the following
resonance parameters of the $\phi(1680)$: $m=(1689\pm 7)$~MeV/$c^2$,
$\Gamma=(211\pm 14)$~MeV/$c^2$ and $\BR(\phi \pp)\times
\BR_{\EE}=(1.24\pm0.09)\times 10^{-7}$, while those of the $\y$ are
$m=(2079\pm 13)$~MeV/$c^2$, $\Gamma=(192\pm 23)$~MeV/$c^2$ and
$\BR(\phi \pp)\times \BR_{\EE} = (1.10\pm 0.10)\times 10^{-7}$,
where the errors are statistical only. We also perform a fit with an
additional incoherent BW function centered near  2.4~GeV/$c^2$. The
fitted parameters of this structure are $m=(2406\pm 32)$~MeV/$c^2$
and $\Gamma=(57\pm 58)$~MeV/$c^2$ with a goodness-of-the-fit of
$\chi^2/ndf=62/52$, corresponding to a C.L. of 15\%, where the
errors are statistical only. The statistical significance of the
structure at 2.4~GeV/$c^2$ is $1.5\sigma$ as determined from the
change in the $\chi^2$ value. A fit with an additional non-resonant
component does not improve the fit quality, and the contribution of
the non-resonant term is negligibly small.

We fit the $\phi \fzero$ cross section distribution with a single BW
function that interferes with a non-resonant component which is
partly from non-$\phi \fzero$ contribution, and partly from the
possible $\phi(1680) \to \phi \fzero$ at the high mass tail of the
$\phi(1680)$, as in BaBar's analysis~\cite{babar2175}. There are two
solutions with very similar resonance parameters. The interference
is constructive for one solution and destructive for the other.
Figure~\ref{belle-oneres}(b) shows the result. The fit yields
$m=(2163\pm 32)$~MeV/$c^2$ and $\Gamma=(125\pm 40)$~MeV/$c^2$, with
a goodness-of-the-fit of $\chi^2/ndf=49/36$, corresponding to a C.L.
of 8.0\%. Here we quote simple averages of the two solutions, and
enlarge the errors to cover the full uncertainties. We also perform
a fit with an additional coherent BW function near
2.4~$\hbox{GeV}/c^2$. The fitted mass and width of the $\y$ are
consistent with the results listed above, and the statistical
significance of the structure at $2.4~\hbox{GeV}/c^2$ is estimated
to be $2.3\sigma$.

Since the differences in the $\y$ resonance parameters from fits to
$\phi\pp$ and $\phi\fzero$ are due to the assumptions on the
background shape and the existence of additional resonances, and
$\phi \fzero$ events are a subsample of $\phi \pp$ events, we take
the values of the $\y$ resonance parameters from the fit to the
$\phi \pp$ cross section with two incoherent resonances as the
central values and the differences are taken as one source of
systematic errors, which is the dominant one for the $\y$ resonance
parameters. For the $\phi(1680)$ and $\y$ resonance parameters, we
have also considered the uncertainties in the absolute mass scale,
the mass resolution, the parameterization of the resonances and
background shape, fit range and possible existence of additional
resonances as systematic errors. Finally, we obtain
$m(\phi(1680))=(1689\pm 7\pm 10)$~MeV/$c^2$,
$\Gamma(\phi(1680))=(211\pm 14\pm 19)$~MeV/$c^2$, and
$\BR(\phi(1680)\to\phi \pp)\times
\BR_{\EE}=(1.24\pm0.09\pm0.14)\times 10^{-7}$, and
$m(\y)=(2079\pm13^{+79}_{-28})$~MeV/$c^2$,
$\Gamma(\y)=(192\pm23^{+25}_{-61})$~MeV/$c^2$, and $\BR(\y\to\phi
\pp)\times \BR_{\EE} = (1.10\pm 0.10\pm0.12)\times
10^{-7}$~\cite{incoherent}, where the first errors are statistical,
the second systematic.


In summary, we present the most precise measurements of the cross
sections for $\EE\to\phi \pp$ and $\EE \to \phi \fzero$ from
threshold to $\sqrt{s}=3.0$~GeV.  The masses and widths of the
$\phi(1680)$ and $\y$ are determined and are in agreement with the
previous measurements~\cite{babar1680,babar2175,bes2175,PDG,DM1}.
The width of the $\y$ tends to be larger than in previous
measurements~\cite{babar2175,bes2175} although the error is large.
We find that the widths of the $\phi(1680)$ and $\y$ are quite
similar and both are at the 200~MeV/$c^2$ level. This may suggest
that the $\y$ is an excited $1^{--}$ $s\bar{s}$ state. Since the
$\fzero$ is thought to have a large $s\bar{s}$ component, $\y\to
\phi\fzero$ can be viewed as an open-flavor decay as opposed to the
case of $Y(4260)\to \jpsi\pp$, which is a hadronic transition. The
study of the $\y$ in other decay modes would be useful for
distinguishing between different possibilities. The branching
fractions of $\jpsi$ decays into $\phi\pp$ and $\phi\fzero$ are
measured with improved precision; the results are in good agreement
with the existing results.

We thank E. Solodov for useful communications. We thank the KEKB
group for the excellent operation of the accelerator, the KEK
cryogenics group for the efficient operation of the solenoid, and
the KEK computer group and the National Institute of Informatics for
valuable computing and SINET3 network support. We acknowledge
support from the Ministry of Education, Culture, Sports, Science,
and Technology (MEXT) of Japan, the Japan Society for the Promotion
of Science (JSPS), and the Tau-Lepton Physics Research Center of
Nagoya University; the Australian Research Council and the
Australian Department of Industry, Innovation, Science and Research;
the National Natural Science Foundation of China under contract
No.~10575109, 10775142, 10875115 and 10825524; the Department of
Science and Technology of India; the BK21 program of the Ministry of
Education of Korea, the CHEP src program and Basic Research program
(grant No. R01-2008-000-10477-0) of the Korea Science and
Engineering Foundation; the Polish Ministry of Science and Higher
Education; the Ministry of Education and Science of the Russian
Federation and the Russian Federal Agency for Atomic Energy; the
Slovenian Research Agency;  the Swiss National Science Foundation;
the National Science Council and the Ministry of Education of
Taiwan; and the U.S.\ Department of Energy. This work is supported
by a Grant-in-Aid from MEXT for Science Research in a Priority Area
("New Development of Flavor Physics"), and from JSPS for Creative
Scientific Research ("Evolution of Tau-lepton Physics").


\begin{thebibliography}{**}
\bibitem{incoherent} We also fit with two coherent BW functions.
There are two solutions with equally good fit quality (the
goodness-of-the-fit is $\chi^2/ndf=64/54$) . The masses and widths
of the resonances are the same for both solutions. The partial
widths to $\EE$ and the relative phases between them are different.
The resonance parameters of the $\phi(1680)$ are $m=(1682\pm
8\pm10)$~MeV/$c^2$, $\Gamma=(226\pm 19\pm19)$~MeV/$c^2$, and
$\BR(\phi \pp)\times \BR_{\EE}=(1.02\pm 0.08 \pm 0.12)\times
10^{-7}$ for the constructive solution, $\BR(\phi \pp)\times
\BR_{\EE}=(1.64\pm 0.08 \pm 0.18)\times 10^{-7}$ for the destructive
solution, and those of the $\y$ are $m=(2097\pm 17
^{+72}_{-45})$~MeV/$c^2$, $\Gamma=(212\pm 31
^{+22}_{-79})$~MeV/$c^2$, and $\BR(\phi \pp)\times
\BR_{\EE}=(0.36\pm 0.07 \pm 0.03)\times 10^{-7}$ for the
constructive solution, $\BR(\phi \pp)\times \BR_{\EE}=(3.82\pm 0.39
\pm 0.38)\times 10^{-7}$ for the destructive solution, where the
first errors are statistical and the second systematic.
\end{thebibliography}
\end{document}